\documentstyle[11pt,paspconf,epsf]{article}

\begin{document}

\title{UVES-VLT Observations of the OIII Bowen Lines in RR Tel}
\author{Pierluigi Selvelli}
\affil{C.N.R.--G.N.A.--
Osservatorio Astronomico di Trieste, Via G.B.Tiepolo 11,
I-34131 Trieste, Italia}
\author {Piercarlo Bonifacio}
\affil{Osservatorio Astronomico di Trieste, Via G.B.Tiepolo 11,
I-34131 Trieste, Italia}

\begin{abstract}
The  exceptional resolution of UVES  has allowed  the
detection  of weak spectral features  and the separation of  
components in blended lines. The intensities  of all of the OIII 
fluorescence lines produced by the O1, O3 and other channels, including the 
5592 \AA ~CE line, have been measured and their ratios compared with models.

\end{abstract}

\section{Introduction}

We have taken advantage of the extremely high spectral resolution in the  
VLT-UVES data to revisit the spectral features
of the  symbiotic nova RR Tel.
We will consider here some  aspects of these new observations, related to 
fluorescence mechanisms in the OIII lines.  
We refer to Selvelli and Bonifacio (2000) for a  detailed description of the 
data and  their reduction.

\section{The OIII Bowen lines}

It is well known  (see Kastner and Bahtia (1996) (KB) for a comprehensive  
paper) that the excitation of the OIII Bowen lines is basically 
due to a fluorescence mechanism in which the He$^+$ resonance  line at 
303.78 \AA, in an accidental near coincidence in wavelength with the  303.80 
\AA ~ resonance transition of O$^{2+}$, pumps the $(2p 3d) ^3P_2$ level of  
O$^{2+}$  (O1 process). Additional fluorescent  processes might be also 
present, i.e., the excitation of the O$^{2+} (2p3d)^3P_1$ level due to 
possible overlap of  the
exciting  He$^+$ line with the 303.69 \AA ~resonance transition of O$^{2+}$
(O3 process), and the (less probable) excitation of the $(2p3d) ^3P_0$ level
by pumping in the resonance line at 303.46 \AA.
Also, a charge-exchange mechanism (CE) seems at
work for some of the lines produced in the decay from the $^3P$ term.
We recall that while some cascade lines are common to 
all processes, there are individual lines  that are "specific", e.g. the
3444 \AA~ and 3132 \AA ~ lines come only  from the O1 process, the 3430 \AA
~ and 3121 \AA ~ only from O3, the 3408 \AA ~ and 3115 \AA ~ only from
pumping of the
303.46 \AA ~ line, and the 5592.34 \AA ~ line (mult.5) from charge-exchange.

 \begin{figure}
\plotone{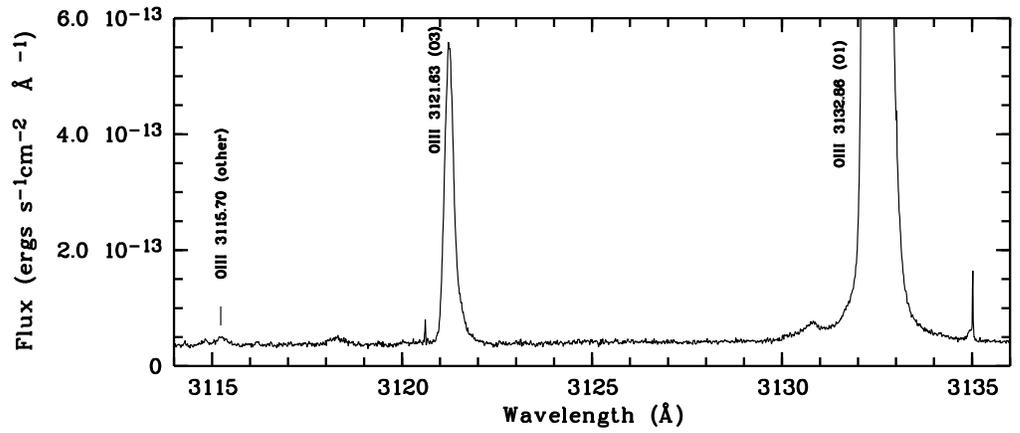}
\caption{A region with OIII lines produced by the single  O1, O3, and
303.46 \AA ~ (other) pumping  processes.} \end{figure}

\begin{table}
\caption{Bowen line intensities in RR Tel.}
\begin{center}
\begin{tabular}{ccc}
\tableline
$\lambda$    & Observed intensities  &  Relative intensities\\
         &  ($10^{-13}$erg cm$^{-2}$s$^{-1}$)  & (I(3444.10)=100)
\\ \tableline
3047.13  &    6.60        &          43.4 \\
3059.30  &    0.67        &           4.4 \\
3115.70  &    0.06        &           0.4 \\
3121.63  &    1.68        &          11.1 \\
3132.86  &   45.50        &         299.3 \\
3299.36  &    1.96        &          12.9 \\
3312.30  &    5.08        &          33.4 \\
3340.74  &    7.17        &          47.2 \\
3405.71  &    0.21 bl.    &        $ <1.4$ \\
3408.11  &    0.21        &           1.4 \\
3415.26  &    0.27        &           1.8 \\
3428.63  &    2.03        &          13.3 \\
3430.60  &    0.31        &           2.0 \\
3444.10  &   15.20        &         100.0 \\
3754.69  &    0.53        &           3.5 \\
3757.24  &    0.10        &           0.7 \\
3759.87  &    blend       &            /  \\
3774.02  &    0.09        &           0.6 \\
3791.27  &    0.156       &           1.0 \\
3810.59  &    blend       &            /  \\
5592.37  &    0.14        &           0.9 \\
H$\beta$ &  101.5         &         667.8 \\
HeII-3203&   25.6         &         168.4 \\
\tableline

\end{tabular}
\end{center}
\end{table}

\begin{figure}
\plotone{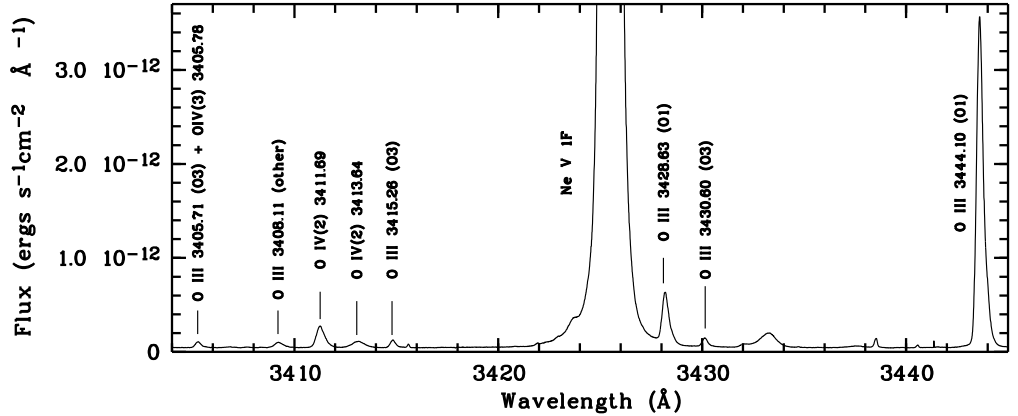}
\caption{Another region with OIII lines produced by the single O1, O3 and
303.46 \AA ~ (other) pumping  processes.}
\end{figure}

The knowledge of the intensities of these lines is fundamental to
determine the relative contribution  to the
observed spectrum  by the various processes previously mentioned. KB have
pointed out that a "complete" Bowen system  has not yet been observed
because of  bad covering in wavelength, scarce spectral resolution, lack of
accurate line intensities, lack of overlap between data of various
observers, the lack of sufficiently accurate  intensities for the weak CE
line at 5592 \AA, and for the O3 process primary cascade line at 3121.6 \AA
~ being particularly disturbing.
 Very recently,
Pereira, de Ara\'ujo, and Landaberry (1999), have presented the results of
new observations in the range 3100-4100 \AA ~on ESO data with spectral
resolution of $\sim 2.5$ \AA ~ FWHM,  in a study of Bowen fluorescence lines
for a group of symbiotic stars, including RR Tel.  The better resolution has
increased the number of lines with adequate measurements and has made
possible a comparison of line ratios with the predictions of models.
However, errors in the line ratios of the order of 10 percent or larger (for
the weaker lines) are still present.

\begin{table}
\caption{Comparison between the observed and
predicted Bowen fluorescence line ratios. }
\begin{center}
\begin{tabular}{ccccc}
\tableline
line ratios &  observed  &   BK   &  FF   &  SS   \\
3133/3444   &  2.99      &  4.06  & 3.21  &  3.61 \\
3428/3444   &  0.13      &  0.18  & 0.15  &  0.34 \\
3405/3411   &  0.78      &  0.75  & 0.78  &   /   \\
3415/3121   &  0.16      &  0.09  & 0.15  &   /   \\
3299/3341   &  0.27      &  0.23  & 0.26  &  0.20  \\
3299/3312   &  0.39      &  0.35  & 0.36  &  0.34  \\
3791/3754   &  0.30      &  0.31  & 0.30  &  0.33  \\
\tableline

\end{tabular}
\end{center}
\end{table}

In our UVES data (spectral resolution $\sim 0.05$ \AA ~ FWHM)
we have detected ALL of the OIII  fluorescent lines in the range
3045-3880 \AA, together with the CE process line at 5592.37 \AA.  Figures 1,
2,  and 3, are self-explaining and illustrate the exceptional quality of
the data that have allowed the detection and measurement of the  "pure"
lines produced in the various excitation process, including the very weak
line at 3408.11 \AA ~ associated with pumping in the 303.46 \AA ~channel .
Table 1 gives the line intensities in erg cm $^{-2}$ s$^{-1}$ and
the ratio with the reference line at 3444.10 \AA. We have
also included for comparison the intensity of H$\beta$ and of the HeII 3203
\AA ~line. The two lines at 3759.87 \AA ~and at 3810.59 \AA ~ are blended
with  stronger lines
(FeVII 3758.92 \AA ~ and OVI(1) 3811.35 \AA, respectively ) and their intensity
 is not
given. The remaining lines are unblended except for the
3405.78 \AA~ line that is partially  affected by the presence of the OIV(3)
line at 3405.78  \AA ~. 
 The high accuracy of the data will allow a detailed  comparison with 
 the predictions of  theoretical models  (Bhatia and Kastner (1993) (BK), 
Froese Fischer (1984) (FF), Saraph and 
Seaton (1980) (SS)). We limit ourselves here, following Pereira, Ara\'ujo,
and Landaberry, to present in Table 2 a comparison between the observed and
the predicted line ratios for some selected lines. It is clear from Table 2
that the best agreement in the overall ratios is with the Froese Fisher
(1994) rates.

\begin{figure}
\plotone{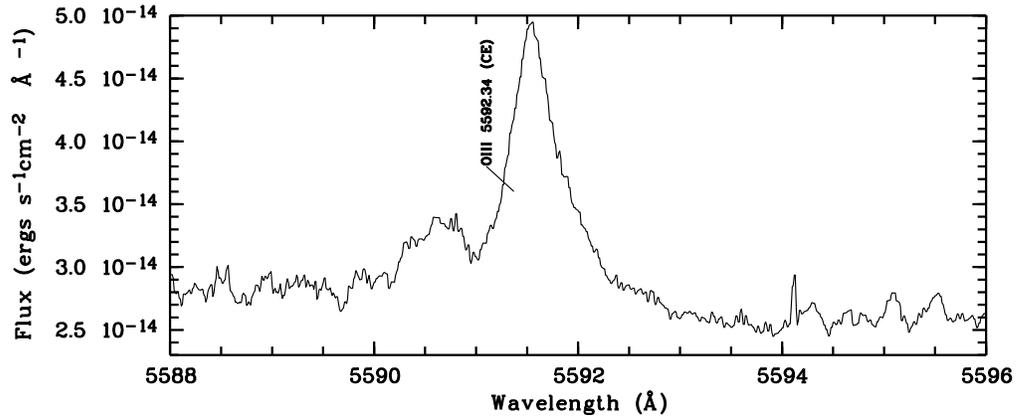}
\caption{The region of the OIII 5592.37 \AA ~ line produced by the
charge-exchange process} \end{figure}

\end{document}